\documentclass[aps,prd,longbibliography,nofootinbib,amsthm,amsmath,amssymb,amsfonts,notitlepage]{revtex4-1}
\usepackage[utf8]{inputenc}
\usepackage[T1]{fontenc}
\usepackage[english]{babel}
\usepackage{graphicx}
\usepackage{float}
\usepackage{cleveref}
\usepackage{xcolor}
\usepackage{color}
\usepackage{tikz}
\usepackage{multirow}
\usepackage{appendix}
\usepackage{slashed}
\usetikzlibrary{arrows}

\date{\today}

\begin{document}
	\title{$Z_n$ symmetry in the vortex muon decay }
	\author{Pengcheng Zhao}
	\email{zhaopch5@mail2.sysu.edu.cn}
	\affiliation{School of Physics and Astronomy, Sun Yat-sen University, 519082 Zhuhai, China}
	
	\begin{abstract}
Polarization of a vortex state fermion has rich structure due to the nontrivial momentum distribution of wave function. This larger freedom provides an unique opportunity to prepare fermions in exotic polarized states, which do not exist for plane-wave state fermions. Based on the so called spin-orbit state which was studied both theoretically and experimentally, we put forward a peculiar vortex muon whose polarization exhibits $Z_n$ symmetry and study its decay. We investigate the  azimuthal distribution of the emitted electrons and find that it exhibits the same symmetry ($Z_n$) as the initial state.

	\end{abstract}
	
	\maketitle
	
	\section{Introduction}
Physics of particles prepared in non-plane-wave states represents a fascinating cross disciplinary topic which attracts much attention \cite{Processes with large impact parameters at colliding beams, Scattering of wave packets with phases}. 
One particularly interesting class of none-plane wave states is vortex beams.
The state carries non-zero orbital angular momentum (OAM) with respect to the propagation direction.
{ Wave functions of the kind of states are always proportional to a spiral phase factor: $\psi(\bold r)=\psi(\rho, \phi, z)\propto \text{exp}(il\phi)$, where $\rho, \phi$ are polar transverse coordinates, $z$ is longitudinal coordinate and $l$ is the integer topological charge. The topological charge acts as an important parameter to determine total angular momentum (TAM) of the state ($\hbar l$ is exactly the OAM for scalar field) and also represents an additional degree of freedom which specifies a peculiar series of non-plane-wave states.}
Optical vortex beams have been studied since the last decade of the 20th century \cite{Screw Dislocations in Light Wavefronts, Astigmatic laser mode
converters and transfer of orbital angular momentum, Orbital angular momentum of
light and the transformation of Laguerre-Gaussian laser modes, Helical-wavefront laser beams
produced with a spiral phaseplate, First Observation of Photons Carrying Orbital
Angular Momentum in Undulator Radiation, Orbital angular momentum beam generation using
a free-electron laser oscillator}, vortex electrons have been produced ten years ago \cite{Generation of electron beams carrying orbital angular momentum}, and recently neutrons \cite{Controlling neutron orbital angular
momentum} and atoms \cite{Vortex beams of atoms and molecules} also have been put into vortex states.
They have already found numerous applications in manipulation of matter, quantum communication, microscopy and so on \cite{Mechanical equivalence of spin and orbital
angular momentum of light: an optical spanner, Dynamic holographic optical tweezers, Microoptomechanical pumps assembled and driven by holographic
optical vortex arrays, Quantum Correlations in Optical Angle-Orbital
Angular Momentum Variables, Terabit free-space data transmission employing
orbital angular momentum multiplexing, Chiral specific electron vortex beam spectroscopy, Probing the
electromagnetic response of dielectric antennas by vortex electron beams, Spin polarisation with electron Bessel beams, Local orbital angular momentum revealed by spiral phase
plate imaging in transmission electron microscopy, Detecting transition radiation from a magnetic moment}.
For fundamental physics, it is interesting to investigate how particles in vortex states will modify standard scattering processes \cite{Colliding particles carrying nonzero orbital angular momentum, Scattering of twisted relativistic electrons by
atoms, Twisted electron
impact single ionization coincidence cross-sections for noble gas atoms, Rutherford scattering of electron vortices, Compton Upconversion of Twisted Photons: Backscattering
of Particles with Non-Planar Wave Functions, Structured x-ray beams from twisted electrons by
inverse Compton scattering of laser light, Elastic scattering of vortex electrons provides
direct access to the Coulomb phase, Radiative Capture of Cold Neutrons by Protons and
Deuteron Photodisintegration with Twisted Beams, Schwinger scattering of twisted neutrons
by nuclei, Delta baryon photoproduction with twisted photons}. 
Recently, the vortex muon decay has been calculated \cite{decay of the vortex muon}.
The spectral-angular distribution of the emitted electron displays several features which are different from plane-wave muons. 

In addition to the OAM, vortex states bring novel polarization opportunities. 
In contrast to plane waves which are described by constant polarization vectors (for photons) or spin vectors (for fermions), polarization of vortex light or fermions can be described with a continuous {\em field} of polarization parameters. 
The much larger freedom of defining a polarization state allows for exotic polarizations which would be unthinkable for plane waves.
{ Particularly, we get TAM eigenstates with rotational-symmetric polarization field.
For these kind of states, the only thing that breaks rotation symmety is the phase distribution, which increases evenly according to the topological charge when we rotate the state by certain axis.
Vortex states with rotational-asymmetric polarization field are not TAM eigenstates and have more complicated structure.}

{A variety of polarized vortex light beams have been generated with a special spatial light modulator accompanied by well-designed optical device series \cite{Method for the generation of arbitrary complex
vector wave fronts, Tailoring of arbitrary optical vector beams}.
The experements ultilize interference of two Laguerre-Gaussian (LG) beams or Hermite-Gaussian (HG) beams with different quantum numbers and different constant polarization modes to acquire light beams with nonconstant polarization modes.
For example, radially polarized beams and azimuthally polarized beams which are rotational-symmetric are generated, and beams named as anti-vortices with peculiar periodic azimuthal polarization distribution which break the rotation symmetry but reserve certain $Z_n$ symmetry are also generated \cite{Tailoring of arbitrary optical vector beams}.}
These kind of states are also called spin-orbit states because it describes states with correlated spin and orbital angular momentum \cite{Spin-orbit states of neutron wave packets}.
Light beams prepared in such states have unusual electromagnetic field distribution and have been suggested to have applications in polarization spectroscopy, linear electron beam accelerator and data storage \cite{Modeling of inverse Cerenkov laser acceleration with axicon laser-beam focusing, Optical guiding of a radially polarized laser beam for inverse Cherenkov acceleration in a plasma channel, Magnetic vortex core reversal by excitation with short bursts of an alternating field}.
{Generation of spin-orbit fermion states have also been proposed for electron and neutron \cite{Generation of a spin-polarized electron beam by multipole
magnetic fields, Methods for preparation and detection of neutron spin-orbit states}.
One method is to propagate an electron or neutron wave packet, which has zero topological charge and constant spin distribution along $z$ direction, through a quadrupole magnetic field whose field intensity vectors locate in one plane and are orthogonal to $z$ direction.
The gererated spin-orbit state is coherent superposition of two states with different topological charges (difference between them can only be one) and opposite polarizations.
It is proposed for electron in \cite{Generation of a spin-polarized electron beam by multipole
magnetic fields} and expanded for neutron in  \cite{Methods for preparation and detection of neutron spin-orbit states}.
Another method is to prepare a state which has zero topological charge and is coherent superposition of both spin up and spin down states, and then propogates it through a magnetic spiral phase plate (mSPP)\cite{Methods for preparation and detection of neutron spin-orbit states}.
If the mSPP is appropriately magnetized, only spin up (or spin down) component changes its topological charge according to the winding number of the mSPP.
By this method, we can get spin-orbit states with arbitrary topological charge components.
It is proposed in \cite{Methods for preparation and detection of neutron spin-orbit states} for neutron and well-designed sequencial quadrupole chain is also proposed to increase range of maximal entanglement.
In principle, both the two methods can be applied to any other fermion as long as its beam is stable enough.
For example, it should be realizable for muon beams with appropriately high energy.}
Other method like using “lattice of optical vortices” prism pairs to put spin-orbit state into lattices is also proposed. Lattices of spin-ortbit neutron state has been claimed to be realized experimentally\cite{Generation and detection of spin-orbit coupled
neutron beams}, but the image is not distinct enough.
Fortunately, similar experiment for light beam gives nice result \cite{Generation of a lattice of spin-orbit beams via coherent averaging}, which shows the feasibility of the method.
{Spin distribution of the anticipant spin-orbit fermion states may also be rotational symmetric (for example, cylindrically polarized states, azimuthally polarized states, radially polarized states, hedgehog skyrmion states and spiral skyrmion states) or exhibit $Z_n$ symmetry (for example, hybrid polarized states and quadrupole spin-orbit states)  as light beams do \cite{Methods for preparation and detection of neutron spin-orbit states}.}

As is realized for photon and suggested for fermions, one can imagine a vortex fermion state, whose local polarization parameters change in the transverse plane as functions of the radial coordinate $r$ and azimuthal coordinate $\varphi$. 
They can exhibit a periodic dependence on $\varphi$ and thus gives vortex states which exhibit $Z_n$ symmetry.
{This kind of state is expected to be generated by the methods discussed above with well-designed experimental installations. }
Instead of continuous azimuthal rotations, such a peculiar fermion state is invariant under its discrete subgroup. 
It represents yet another adjustable degree of freedom whose physical significance and potential application remain essentially unknown up to now.
In this paper, we explore some properties of the new degree of freedom, which is specialized by positive integer number $n$ from $Z_n$ symmetry of the state, in a simple setting: decay of the vortex muon prepared in a polarization state with $Z_n$ symmetry. 
Our previous work on vortex muon decay \cite{decay of the vortex muon} assumed the polarization state to be invariant under continuous azimuthal rotations and it resulted in electron distribution that is also rotational symmetric. 
{In that paper, we detailedly discussed kinematical characteristics of vortex muon decay.}
Now, we explore the case that the vortex muon state exhibits $Z_n$ symmetry and discuss whether the $Z_n$ symmetry will remain for the final state after decay of the polarized vortex muon by exploring angular characteristics.

The paper is organized as follows.
In section \ref{section two}, we describe vortex states with $Z_n$ symmetry.
In section \ref{section three}, we give general formulas for calculation of vortex muon decay at tree level.
In section \ref{section four}, we prove that the initial state and emitted electron distribution for polarized vortex muon decay exhibits the same $Z_n$ symmetry and give some examples.
In section \ref{section five}, we draw the conclusions.
In this paper, the natural units are used.
Average 3-momentum direction of initial state is chosen to be $z$ axis.
We use symbol $\vec r$ to represent transverse 2D vectors,  $\bf r$ to represent 3D vectors and $r^{\mu}$ to represent 4D vectors.

\section{Vortex muon state with $Z_n$ symmetry}\label{section two}
\subsection{From plane wave state to vortex state}
As a fermion, single muon state is given by the solution of the Dirac equation.
In the standard representation, the plane wave muon state in coordinate space is:
\begin{equation}\label{plan wave spinor}
\psi^{PW}(x)=u(p^{\mu},s^{\nu})e^{-iEt+i\bold p \cdot \bold x},
\end{equation}
where $p^{\mu}=(E,\bold p)$ is the 4-momentum of muon, $E$ is muon energy, $\bold p$ is muon 3-momentum,
$s^{\nu}$ represents spin of muon.
The spinor can be written as
\begin{equation}\label{spinor}
u(p^{\mu},s^{\nu})=\frac{1}{\sqrt{2E}}\left(\begin{array}{l}
\sqrt{E+m} \,w \\
\sqrt{E-m} \,\bold{\sigma} \cdot \bold n \,w
\end{array}\right), \quad w=\left(\begin{array}{l}
\cos \frac{\alpha}{2} \\
\sin \frac{\alpha}{2} e^{i \delta}
\end{array}\right).
\end{equation}
$m$ is muon mass, $\bold n$ is unit vector along muon's direction of motion.
Polarization of the state is described by the spinor with two free parameters $\alpha$ and $\delta$ or 4-vector $s^{\nu}$.
There is one-to-one correspondence between them.
Polarization 4-vector of completely polarized muon is $s^{\nu}=(S^0,\bold S)=(\gamma \beta (\bold n \cdot \bold s),\bold s+(\gamma-1)(\bold n \cdot \bold s)\bold n )$, where $\bold s$ is unit vector in muon spin direction in the rest frame, $\beta$ is muon speed and $\gamma=1/\sqrt{1-\beta ^2}$ \cite{quantum electrodynamics}.

\begin{figure}[!h]
\centering
\includegraphics[width=0.6\textwidth]
{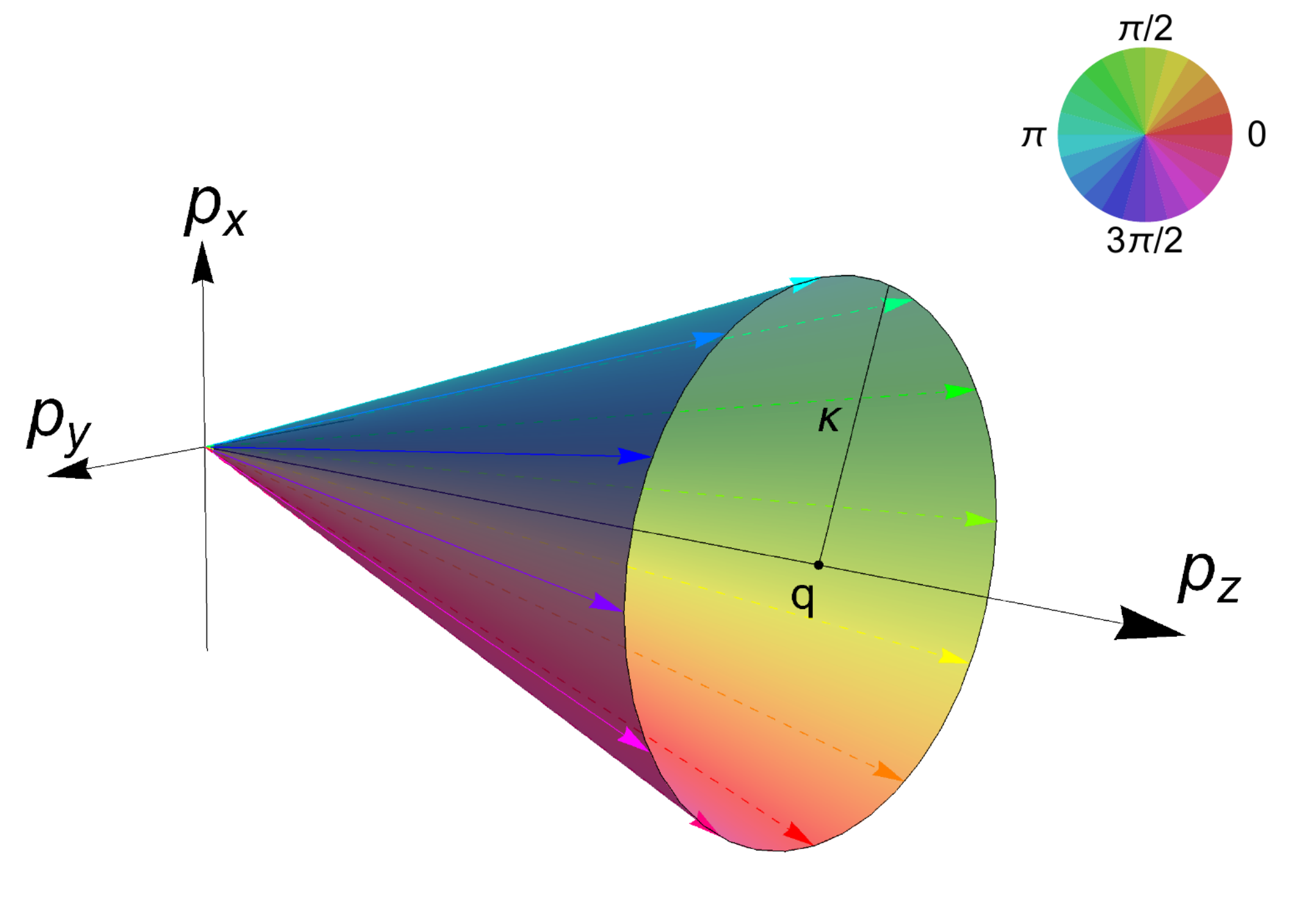}
\caption{Bessel vortex state with $l=1$. It is a monochromatic state. The black circle with radius $\kappa $ gives distribution in momentum space. All the momentum vectors form a circular cone. Phase factors of plane wave components changes $2\pi l$ in one circle (different phases are shown by different colors) and $l$ acts as vortex quantum number (topological charge) of the state. $q$ is longth of average momentum. $\kappa$ is longth of transverse momentum for all plane wave components, which also acts as transverse quantum number of the state.}\label{bessel vortex state}
\end{figure}

As for the vortex state, it can be described by linear combination of many different plane waves.
{The most simple vortex state is the Bessel vortex state.
The wave function of Bessel fermion state is written as \cite{Generation of High-Energy Photons with Large Orbital Angular Momentum
by Compton Backscattering, Colliding particles carrying nonzero orbital angular momentum}
\begin{eqnarray}\label{coordinate space state}
\psi^V_{E,\kappa,l}(x)&=&e^{-iEt+iqz}\int \frac{d^2\vec k}{(2\pi)^2}a_{\kappa l}(\vec k)u(p^{\mu},\,s^{\nu})e^{i\vec k \cdot \vec {\rho}}\nonumber\\
a_{\kappa l}(\vec k)&=&(-i)^l\sqrt {\frac{2\pi}{\kappa}}\delta(|\vec k|-\kappa)e^{il\varphi_p},
\end{eqnarray}
where $p^{\mu}=(E,\,\vec k,\,q)$ is 4-momentum of each plane wave component, $E$ is energy, $\vec k=(|\vec k|,\,\varphi_p)$ is the transverse 2-dimentional projection of $p^{\mu}$, $q$ is longitudinal momentum, $x^{\mu}=(t,\,\vec {\rho},\,z)$ is the 4-vector in coordinate space,  $t$ is time, $\vec {\rho}=(|\vec {\rho}|,\,\varphi_r)$ is the transverse 2-dimentional projection of $x^{\mu}$, $z$ is longitudinal space coordinate, $\kappa$ is transverse momenta modulus, $l$ is topological charge (integer quantum number for vortex) of Bessel type vortex state.
In the momentum space, its distribution forms a circle that is orthogonal to propogation direction and all the momentum vectors form a circular cone as Fig. \ref{bessel vortex state} shows.
Three quantum numbers $E$, $\kappa$, and $l$ uniquely determine the state.
Other vortex state like LG state can be regarded as superposition of many Bessel type vortex states.
Since its simplicity, we just discuss Bessel type vortex state in this paper.}

Note that, two free parameters $\alpha$ and $\delta$ in the spinor $u(p^{\mu },s^{\nu })$ for different plane wave components can be the same or different.
For the case that they keep the same, the state must be rotational-symmetric\cite{decay of the vortex muon, Relativistic Electron Vortex Beams: Angular Momentum and Spin-Orbit Interaction, Scattering of twisted relativistic electrons by atoms
}.
For the case that we have angular dependent parameters $\alpha (\varphi_p)$ and $\delta(\varphi_p)$, the state can have diverse structures.
It's these dependence that indicate the possibility of constructing vortex state with $Z_n$ symmetry.

\begin{figure}[!h]
		\centering
\includegraphics[width=\textwidth]{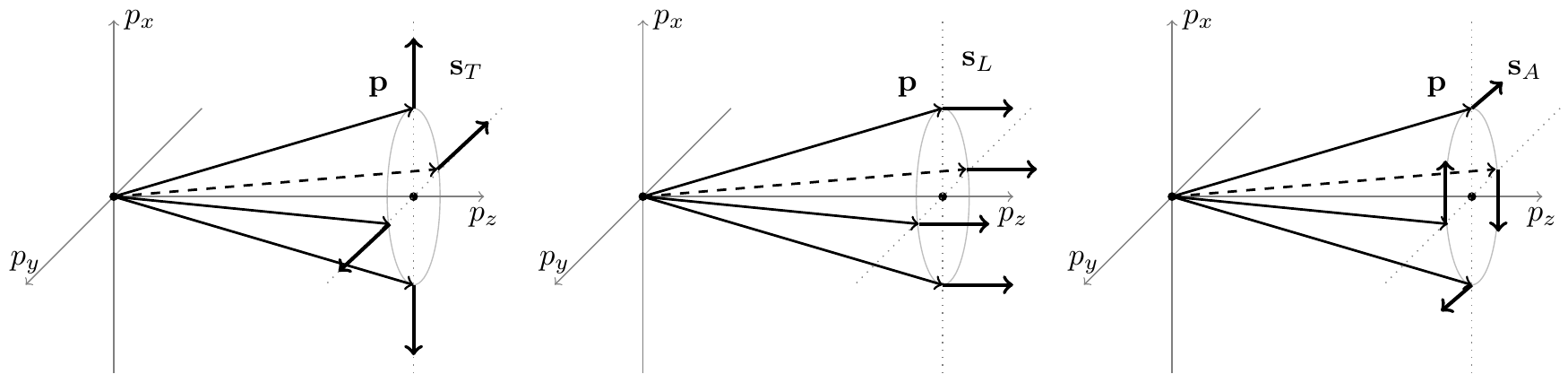}
		\caption{Three classes of basic vectors in momentum space for describing  polarization of vortex particles: transverse basic vectors (left), longitudinal basic vectors (center), and azimuthal basic vectors (right).
Every plane wave component has polarization vector $\bold s$ that is combined by three orthogonal basic vectors from different classes.}\label{Fig-polarization-states}
	\end{figure}

Since $\bold S$ is a three-dimensional vector, it is convenient to choose three basic vectors to represent it.
The three unit vectors for every plane wave component with momentum $\bold p$ are selected as $(\pi/2,\varphi_p)$ (transverse), $(0,0)$ (longitudinal) and $(\pi/2,\varphi_p-\pi/2)$ (azimuthal), which are expressed in the spherical coordinate.
They are visualized in Fig.\ref{Fig-polarization-states} and labeled as $\bold s_{T},\bold s_{L}, \bold s_{A}$ respectively.
	Generally, for arbitrary polarized vortex state, every plane wave component in it has its own 4-vector $s^{\nu}=(S^0,\bold S)$: 
\begin{eqnarray}\label{jihuafenjie}
&&S^0=\gamma \beta\,(\bold n \cdot \bold s)=\beta\,(\bold n \cdot \bold S),\nonumber\\
&&\bold S=a_p\,\bold s_{T}+b_p\,\bold s_{L}+c_p\,\bold s_{A},
\end{eqnarray}
	where $ a_p,\,b_p,\,c_p$ are functions of $\varphi_p$.
	{They satisfy normalization condition (seeing Appendix A)
\begin{equation}\label{abceq}
(a_p\,\sin{\theta_0}+b_p\,\cos{\theta_0})^2+\gamma ^2\,(a_p\,\cos{\theta_0}-b_p\,\sin{\theta_0})^2+\gamma ^2\,c_p^2=\gamma ^2, 
\end{equation}
where $\theta_0=\arctan {(\kappa /q)}$ is conical angle of vortex muon in momentum space.
This equation means that the three parameters $a_p, b_p, c_p$ are not independent and it is coincident with the fact that plane wave muon has two degrees of spin freedom.}

Polarization of plane wave muon is described by one constant vector.
By combination of different plane waves with different momenta and polarizations, the spin distribution of none-plane wave fermion can have complicated structure.
Polarized vortex muon state can be designed with azimuthal-periodic structure, which exhibits $Z_n$ symmetry.
This is what we call ``vortex muon state with $Z_n$ symmetry'', which corresponds to the case that parameters $a_p,\,b_p,\,c_p$ behave as periodic functions of $\varphi_p$ with period $2\pi /n$ ($n$ can be any positive integer).
Integer n of $Z_n$ symmetry is another peculiar number that specialize the muon state except for muon energy $E$, transverse quantum number $\kappa$ and topological charge $l$.

\subsection{Visualization of vortex muon with $Z_n$ symmetry}
To visualize the polarized vortex photon, we can use the classical electrodynamics quantities: the electric and magnetic field.
But there is no such kind of classical quantity for fermions.
In quantum mechanics, we describe the polarization of a plane wave fermion with spin.
Under non-relativistic limitation, we can use spin field to visualize the polarization of vortex muon with $Z_n$ symmetry.

To give a simple example, we apply paraxial approximation, which means the conical angle $\theta_0$ is much small.
Then the monochromatic muon states we discuss can approximately be eigenstate of OAM and spin operator simultaneously.
Here, we just use this approximation to facilitate the visualization of the spin field of our state.
Later calculation about muon decay is not limited with it.

It is convenient to set $a_p,\,b_p$ and $c_p$ instead of $\alpha (\varphi_p)$ and $\delta (\varphi _p)$ as periodic functions of $\varphi_p$ to construct the kind of vortex muon we need.
Relations between the two groups of parameters are given in Appendix A.
Particularly, with non-relativistic limitation and the paraxial approximation ($\sin \theta_0 \rightarrow 0$ and $\cos \theta _0 \rightarrow 1$), we choose these special cases:
\begin{equation}
b_p=\gamma\,b,\quad a_p=\sqrt{1-b^2}\cos{n\varphi_p},\quad c_p=\sqrt{1-b^2}\sin{n\varphi_p},
\end{equation}
with $-1<b<1$ and $n$ being a positive integer.
Then integral of the state in Eq.\eqref{coordinate space state} gives
\begin{equation}
\psi^V(x)=\sqrt{\frac{\kappa}{2\pi}}
\left(\begin{array}{l}
\sqrt{\frac{b+1}{2}}J_l(\kappa \rho)\\
\sqrt{\frac{1-b}{2}}J_{l+1-n}(\kappa \rho)e^{i[(1-n)\varphi_r]}
\end{array}\right)e^{il\varphi_r}\,e^{-iEt+iqz},
\end{equation}
where $J_l$ is Bessel functions of first kind with order $l$.
The other two componets of spinor $u(p^{\mu},\,s^{\nu})$ tend to $0$ due to non-relativistic limitation and are ignored.
Thus, we can describe the spin distribution in coordinate space with up and down type spinors
\begin{eqnarray}\label{up and down}
\psi^V(x)&=&P_{up}(\rho,\,\varphi_r,\,z)\,|\!\uparrow \rangle+P_{down}(\rho,\,\varphi_r,\,z)\,|\!\downarrow \rangle\nonumber\\
&=&\frac{1}{N}\left( \sqrt{\frac{b+1}{2}}J_l(\kappa \rho)\,|\!\uparrow \rangle+\sqrt{\frac{1-b}{2}}J_{l+1-n}(\kappa \rho)e^{i[(1-n)\varphi_r]}\,|\!\downarrow \rangle \right)\,e^{il\varphi_r}\,e^{-iEt+iqz} ,
\end{eqnarray}
where
\[N=\sqrt{\frac{b+1}{2}J_l^2(\kappa \rho)+\frac{1-b}{2}J_{l+1-n}^2(\kappa \rho)}\]
is $\rho$-dependent coefficient.
The spin field can be described by arrows filling all the space.
The arrows are drawn by method of ``Poincare sphere'' according to difference between $P_{up}$ and $P_{down}$\cite{Structured light}.

\begin{figure}[!h]
	\centering
	\includegraphics[width=0.48\textwidth]{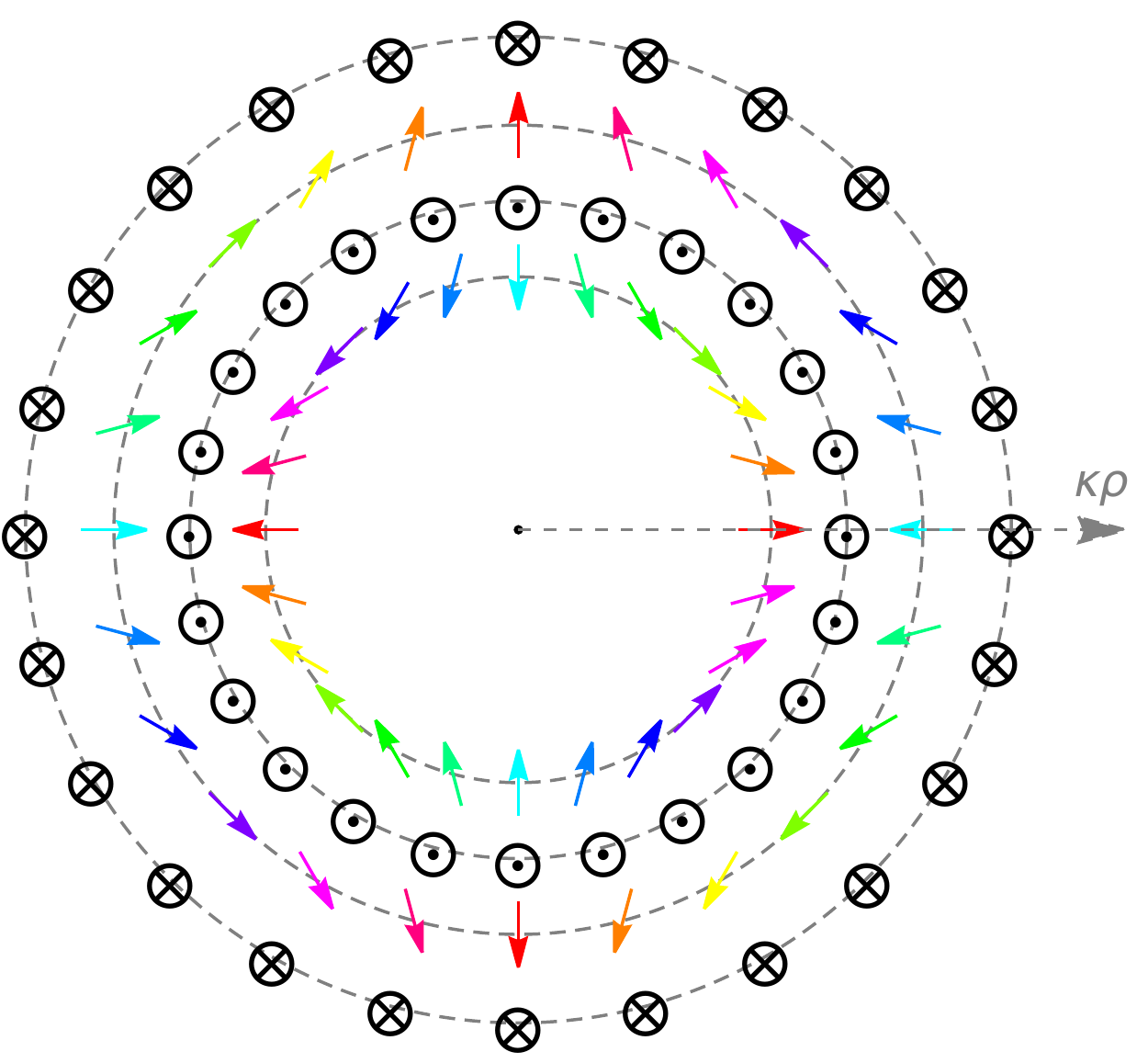}\hfill
	\includegraphics[width=0.48\textwidth]{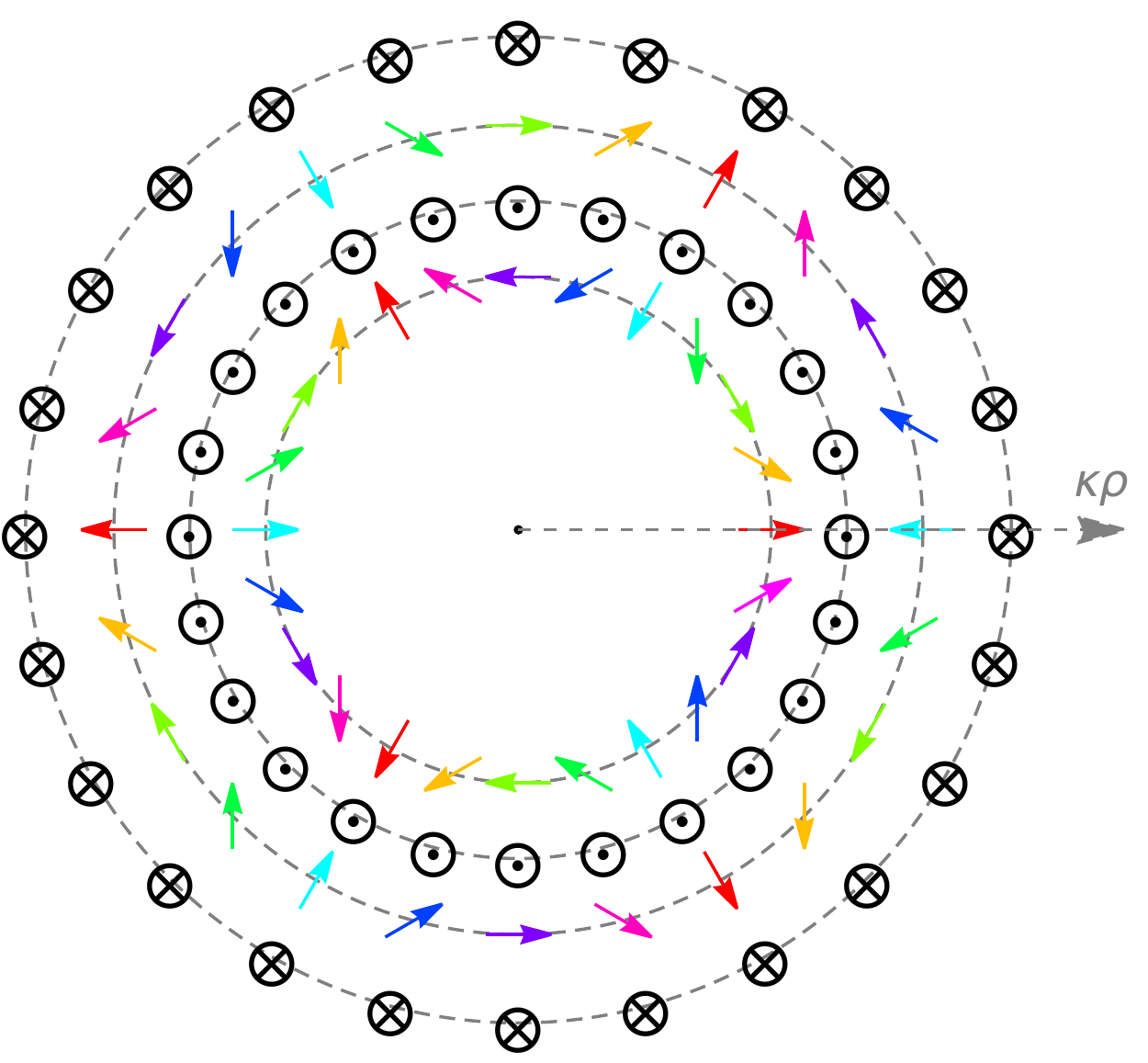}
	\caption{Coordinate-space spin distribution of state described in Eq. \eqref{up and down} in transverse plane in case that $b=0$ and $l=12$. $n=2$ for left picture and $n=3$ for right one. The arrows represent spin at every space point. Only arrows that are completely transverse or longitudinal are drawn. Radial length is not well-proportioned here. Different colors correspond to different direction in transverse plane and show clearly the $Z_n$ symmetry ($Z_2$ for left side and $Z_3$ for right side). ``$\odot$'' and ``$\otimes$'' mean spin to $z$ and opposite drections respectively.}\label{spin distribution}
\end{figure}

For a spesific case, we set $b=0$.
This gives the state with polarization 3-vector $\bold S$ of every plane wave component all locating in the transverse plane in the momentum space.
In the coordinate space, the spin distribution can be complicated.
Its stucture is fully determined by number $l$, $\kappa$ and $n$.
One example with $l=12$ and $n=3$ is showed in Fig.\ref{spin distribution}.
In the figure, we show the special concentric circles on which the arrows that represent spin locate in transverse plane or are orthoganal to the plane.
They correspond to the cases that the two Bessel functions in Eq.\eqref{up and down} have the same modulus or one of the Bessel functions is equal to zero.
In principle, there are an ocean of concentric circles.
We draw four smallest ones in the figure.
The center of the circles is singularity of vortex state.
At points in the empty space, either in the smallest circle or between adjacent concentric circles, the arrows have both transeverse and langitudinal components.
The most notable feature of the figure is that the spin distribution is rotational-invariant about $z$ axis under certain rotation angles, i.e. integer multiples of $2\pi /n$.
This kind of symmetry is just $Z_n$ symmetry we designed for our vortex state in momentum space.

\section{Muon decay from plane wave case to vortex case}\label{section three}
To simplify the calculation, electron and neutrino masses are set to zero, and only tree level decay process is considered.
Either the plane wave muon or the vortex muon that we consider are completely polarized.
We review the plane wave muon decay first and then give formulas for the vortex muon decay.

	\subsection{Decay of the plane wave muon}
The symbols we use here are showed as follows.
The two constants $m$ and $G_F$ are muon mass and Fermi constant.
The unit vector of $z$ direction is $\bold n_z$.
$p^{\mu}$ and $k^{\mu}$ are four-momentums of muon and electron.
We write $p^{\mu}=E(1,\beta \bold n)$, where $E,\,\beta,\,\bold n$ are muon energy, muon speed and unit vector of muon direction of motion respectively.
We write $k^{\mu}=E_e(1,\bold n_e)$ with $\bold n_e \cdot \bold n_z=\cos {\theta_e}$, where $E_e,\,\bold n_e,\,\theta_e$ are electron energy, unit vector of electron direction of motion and electron emission angle respectively.
Electron moving direction is $(\theta_e,\phi_e)$.

	{The spectral-anglular distribution for vortex muon is written as \cite{decay of the vortex muon}}
	\begin{eqnarray}\label{plane wave case}
			\frac{d\Gamma_{PW}}{dE_e\, d\Omega}&=&\frac{G_F^2E_e}{48\pi^4 E}\Bigl\{(pk)(3m^2 - 4(pk)) - m(sk)(m^2 - 4(pk))\Bigr\}\nonumber\\
&=&\frac{G_F^2E_e}{48\pi^4 E}\Bigl\{[3 m^2 E E_e(1-\beta\, \bold n \cdot \bold n_e) - 4 E^2 E_e^2 (1-\beta\, \bold n \cdot \bold n_e)^2] - mE_e(S^0-\bold S \cdot \bold n_e)
[m^2 - 4 E E_e (1-\beta\, \bold n \cdot \bold n_e)]\Bigr\}\,.
	\end{eqnarray}
	The formula in curly brackets is a sum of two different terms.
The first term corresponds to the unpolarized plane wave muon case.
	The second term corresponds to the modification of polarized case to unpolarized case.
	What should be noted is that the maximum energy of the emitted electron depends on the emission angle $\theta_e$:
	\begin{equation}
		E_{em} = \frac{m^2}{2E (1-\beta\, \bold n \cdot \bold n_e)}\,.\label{Emax}
	\end{equation}

According to Eq. \eqref{plane wave case}, we can get electon spectral-azimuthal distribution as function of $E_e$ and $\phi_e$ at fixed $\theta_e$.
It will be axisymmetric around $z$ axis, i.e. independent of azimuthal angle $\phi_e$, if both $\bold n$ and $\bold S$ are parallel to $z$ axis.
Or else, it will be $\phi_e$-dependent.
If $\bold n$ is parallel to $z$ axis while $\bold S$ is not, the dependence will be described by trigonometric functions of $\phi_e$ with frequency $1$, i.e. $\cos \phi_e$ and $\sin \phi_e$.
If $\bold n$ is not parallel to $z$ axis, the dependence will be described by trigonometric functions with both $1$ and $2$ frequencies, i.e. $\cos \phi_e$, $\sin \phi_e$, $\cos 2\phi_e$ and $\sin 2\phi_e$.
Note that the axisymmetric contribution ({\em {i.e.}} constant term) always exist.

	\subsection{Decay of the vortex muon}
Bessel type vortex muon is combination of equal-weight plane waves with azimuthally distributed phase factors.
Its differential decay width is angular average of decay width of every plane wave components:
	\begin{equation}
		d{\Gamma_V} = \int \frac{d\varphi_p}{2\pi}\, d\Gamma_{PW}(\bold p)\,.\label{vortex-decay-1}
	\end{equation}

	If $\theta_e$ and $\theta_0$ are fixed, the plane wave components' contribution to the spectral-angular distribution is classified into two classes by two characteristic electron energy: $E_{e1}$ and $E_{e2}$ \cite{decay of the vortex muon}.
Electrons with energy smaller than $E_{e1}$ can be produced by any plane wave components, but electrons with energy bigger than $E_{e1}$ and smaller than $E_{e2}$ can only be produced by sectional components whose azimuthal angle $\varphi_p$ runs over $(\phi_e-\tau,\,\phi_e+\tau)$.
	The two energies are expressed as
	\begin{equation}
		E_{e1} = \frac{m^2}{2E (1-\beta\cos(\theta+\theta_0))}\quad \mbox{and} \quad
		E_{e2} = \frac{m^2}{2E (1-\beta\cos(\theta-\theta_0))}\,.\label{Emax12}
	\end{equation}
	And the critical angle $\tau$ satisfies
	\begin{equation}
		\cos\tau = \frac{(E_e-E_{e1})E_{e2} - (E_{e2}-E_e)E_{e1}}{E_e(E_{e2}-E_{e1})}\,. \label{tau}
	\end{equation}
	
	The spectral-angular distribution for polarized vortex muon decay is written as \cite{decay of the vortex muon}
	\begin{equation}\label{vortex muon decay}
\displaystyle\frac{d\Gamma_{V}}{dE_e\, d\Omega}= \frac{G_F^2E_e}{48\pi^4 E}\int_{\phi_e-\tau}^{\phi_e+\tau}\frac{d\varphi_p}{2\pi} \Bigl\{[3 m^2 E E_e(1-\beta\, \bold n \cdot \bold n_e) - 4 E^2 E_e^2 (1-\beta\, \bold n \cdot \bold n_e)^2]- mE_e(S^0-\bold S \cdot \bold n_e)
 [m^2 - 4 E E_e (1-\beta\, \bold n \cdot \bold n_e)]\Bigr\}\,.
	\end{equation}
	For $E_{e1}<E_e<E_{e2}$, we have $0<\tau<\pi$, while for $0<E_e<E_{e1}$, just set $\tau \rightarrow \pi$.

We take modification of polarized case to unpolarized case out of Eq.~\eqref{vortex muon decay}:
	\begin{equation}\label{vortex muon decay modification}
		\frac{d\Gamma_{V}^{mod}}{dE_e\, d\Omega}= \displaystyle\frac{G_F^2E_e}{48\pi^4 E}\int_{\phi_e-\tau}^{\phi_e+\tau}\frac{d\varphi_p}{2\pi} \Bigl\{- mE_e(S^0-\bold S \cdot \bold n_e)[m^2 - 4 E E_e (1-\beta\, \bold n \cdot \bold n_e)]\Bigr\}\,.
	\end{equation}

	The direction of unit vectors $\bold n$ and $\bold n_e$ are respectively $(\theta_0,\varphi_p)$ and $(\theta_e,\phi_e)$.
Product of unit vectors in direction $(\theta_1,\varphi_1)$ and $(\theta_2,\varphi_2)$ is $``\cos \theta_1\cos \theta_2+\sin \theta_1\sin \theta_2\cos (\varphi_1-\varphi_2)"$.
Then we get
	\begin{equation}\label{vortex muon decay modification 2}
		\frac{d\Gamma_{V}^{mod}}{dE_e\, d\Omega}= \frac{G_F^2mE_e^2}{48\pi^4 E}\int_{\phi_e-\tau}^{\phi_e+\tau}\frac{d\varphi_p}{2\pi} \Bigl\{[\beta (a_p\sin\theta_0+b_p\cos\theta_0)- a_p(\bold s_{T} \cdot \bold n_e)-b_p(\bold s_{L} \cdot \bold n_e)-c_p(\bold s_{A} \cdot \bold n_e)] [ 4 E E_e (1-\beta\, \bold n \cdot \bold n_e)-m^2]\Bigr\}\,,
	\end{equation}
	where
	\begin{eqnarray}
		&&\bold s_{T} \cdot \bold n_e=\sin \theta_e \cos (\varphi_p-\phi_e),\nonumber\\
		&&\bold s_{L} \cdot \bold n_e=\cos \theta_e,\nonumber\\
		&&\bold s_{A} \cdot \bold n_e=\sin\theta_e \sin(\varphi_p-\phi_e),\nonumber\\
		&&\bold n \cdot \bold n_e=\cos \theta_e \cos \theta_0+\sin \theta_e \sin \theta_0\cos (\varphi_p-\phi_e).\nonumber
	\end{eqnarray}

	\section{Electron distribution with $Z_n$ symmetry after muon decay}\label{section four}
\begin{figure}[!h]
		\centering
		\includegraphics[width=\textwidth]{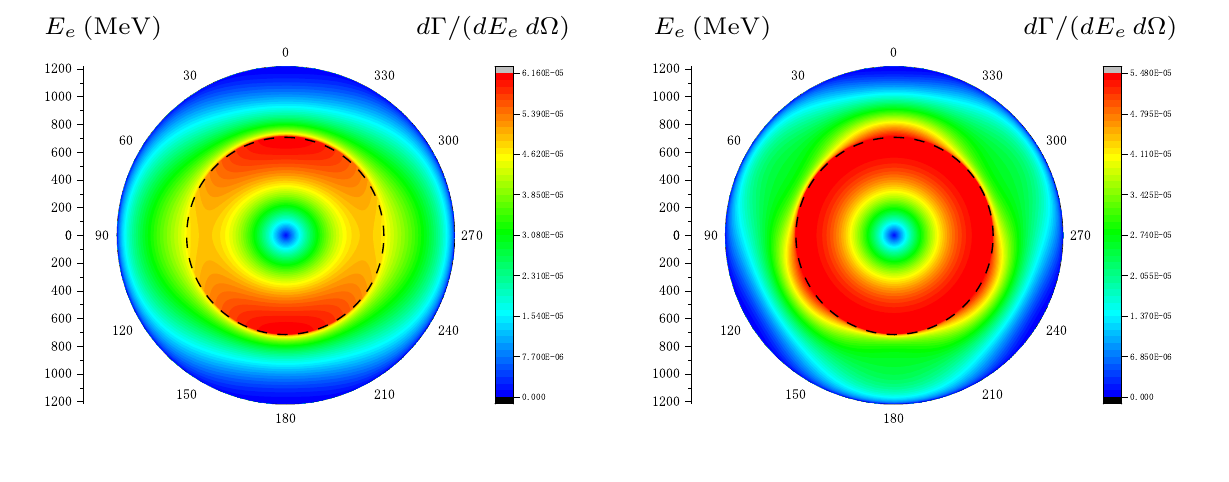}
\caption{Contour map of spectral-azimuthal distribution at $\theta_e=\pi/60$ for emitted electron. The figures are drawn in polar coordinate system. Polar angle represents electron azimuthal angle $\phi_e$ and radial axis represents electon energy $E_e$. 
The contours with different colors represent dimentionless differential decay width ${d\Gamma _V}/{(dE_e\,d\Omega)}$ of muon decay.
The black dashed circles correspond to cases of $E_e=E_{e1}$.
The initial muon state is fixed by parameters: $E=3.1\,\rm{GeV}$, $\theta_0=0.01$, $l=12$, $a_p={\gamma \,\cos {n\varphi_p}}/{\sqrt {\gamma ^2\,\cos^2\theta_0+\sin^2\theta_0}}$, $b_p=0$ and $c_p=\sin n\varphi_p$ (Equation \eqref{abceq} is satisfied).
The state exhibits $Z_n$ symmetry.
$n=2$ for the left picture and $n=3$ for the right picture.
}\label{electron azimuthal distribution z}
	\end{figure}

We reconstruct the integral in Eq. \eqref{vortex muon decay modification 2} as 
\begin{equation}\label{cos and sin integral}
		\frac{d\Gamma_{V}^{mod}}{ dE_e \,d\Omega}= \displaystyle\frac{G_F^2mE_e^2}{48\pi^4 E}\, \int_{\phi_e-\tau}^{\phi_e+\tau}\frac{d\varphi_p}{2\pi} \Bigl\{a_p \sum_{f=0}^{2}\\x_f\cos[f(\varphi_p-\phi_e)]+b_p \sum_{g=0}^{1}\\y_g\cos[g(\varphi_p-\phi_e)]+c_p \sum_{h=1}^{2}z_h\sin[h(\varphi_p-\phi_e)]\Bigr\}\,.
	\end{equation}
Here $x_f,\,y_g$ and $z_h$ are all independent of $\varphi_p$ and $\phi_e$.
They are determined by parameters of the initial state and emitted electron.
As the vortex muon state we consider exhibits $Z_n$ symmetry, $a_p,\,b_p$ and $c_p$ must be periodic functions of $\varphi_p$ with period $2\pi/n\,(n=1,2,3...)$.
Thus they can be expanded as trigonometric series:
\[\frac{1}{2} A_{0}+\sum_{\zeta=1}^{\infty}\left[ A_{n} \cos (\zeta\,n \varphi_p)+B_{n} \sin (\zeta\,n \varphi_p)\right]. \]
We will see later that this expansion extremely simplifies our discussion.

First, we look into the electrons with energy $0<E_e<E_{e1}$, where all plane wave components contribute to the result and we set $\tau=\pi$ in Eq. \eqref{cos and sin integral}. According to the orthogonality of trigonometric integration, we can get the result that the integrals of $``\cos(\varphi_p-\phi_e)"$, $``\cos[2(\varphi_p-\phi_e)]"$, $``\sin(\varphi_p-\phi_e)"$ and $``\sin[2(\varphi_p-\phi_e)]"$ in Eq. \eqref{cos and sin integral} work as filters to keep the same trigonometric term and filter all the other terms in periodic functions $a_p,\,b_p,\,c_p$.
And reserved azimuthal distribution behaves with $``\cos(\phi_e)"$, $``\cos(2\phi_e)"$, $``\sin(\phi_e)"$ and $``\sin(2\phi_e)"$.
The terms independent of $\phi_e$ give constant azimuthal distribution.
The final azimuthal distribution can get contribution from trigonometric functions with frequency $1$ or $2$ but no contribution from ones with bigger frequencies.
It can only have $Z_2$ symmetry or no $Z_n$ symmetry, which both come from symmetry of initial state.
$Z_n$ symmetries with $n\geqslant 3$ can not be delivered to final state in this energy region.

Second, we look into the electrons with energy $E_{e1}<E_e<E_{e2}$, where only sectional plane wave components contribute to the result and we keep the formula of $\tau$ in Eq. \eqref{cos and sin integral}.
The $\tau$, which satisfies Eq. \eqref{tau}, is a constant when initial muon and final electron state are fixed.
This simplifies the integral.
We insert integrals listed in Appendix B into Eq. \eqref{cos and sin integral}.
And then we can see that the integral with $\cos(\varphi_p-\phi_e)$ and $\cos[2(\varphi_p-\phi_e)]$ work as amplitude modulators to rescale the amplitudes in front of trigonometric series with different frequency, while the integral with $\sin(\varphi_p-\phi_e)$ and $\sin[2(\varphi_p-\phi_e)]$ work as amplitude and phase united modulators to rescale the amplitudes and change the modes between $\cos{(\zeta n\phi_e)}$ and $\sin{(\zeta n\phi_e)}$ at the same time.
The characteristic of these integrals causes the final result of Eq. \eqref{cos and sin integral} keeping the frequencies of every trigonometric components in $a_p(\varphi_p),\,b_p(\varphi_p),\,c_p(\varphi_p)$, and thus the spectral-azimuthal distribution (as functions of $\phi_e$) also shows $Z_n$ symmetry as initial vortex muon state does for any positive integer number $n$.

Specially, examples are showed in Fig. \ref{electron azimuthal distribution z}.
A black dashed circle at $E_e=E_{e1}$ is drawn for distinguishing of the two different energy regions in both pictures.
In the left picture, the initial state exhibits $Z_2$ symmetry and we can see that the electron distribution also exhibits $Z_2$ symmetry no matter in the circle or outside the circle.
This is consistent with our discussion above for $n\leqslant 2$.
In the right picture, the initial state exhibits $Z_3$ symmetry and we can see that the electron distribution  exhibits $Z_3$ symmetry outside the circle but is rotationally symmetric in the circle.
This is also consistent with our discussion above for $n\geqslant 3$.
In a global view, we can conclude that final electron distribution have the same $Z_n$ symmetry with initial muon state for any positive  integer $n$.
And if initial state has no $Z_n$ symmetry, the electron distribution will also has no $Z_n$ symmetry.

Note that there is possibility that initial state with $Z_n$ symmetry may results in azimuthal distribution of electron (in the cases that energy $E_e$ and emission angle $\theta_e$ are fixed) with $Z_{n^{\prime}}$ symmetry, where $n^{\prime}$ is integer multiple of $n$.
But this is not a general result.
It just corresponds to the case that the $\cos(n\phi_e)$ and $\sin(n\phi_e)$ terms disappear together for coincidental electron energy and emission angle.
But it does not influence the final result, because all the possible electron energy and emission angle should be read in conjunction.
Global electron azimuthal distribution can only exhibit the same $Z_n$ symmetry as the initial state does.

The result is not difficult to be understood. To do it, we need to keep three prerequisites in mind:
\begin{itemize}
\item
{ There is no interference between any plane wave components in vortex state.}
Square of $M$ matrix for vortex muon decay can be deduced as angular average of $M^2$ for all plane wave components.
This applies to all processes with only one vortex initial state \cite{Colliding particles carrying nonzero orbital angular momentum}.
Thus we can think of vortex muon decay by simply imagining superposition of decay of infinite plane wave muons whose momentums and polarizations are determined by vortex state.
\item
{ For the plane wave muon decay, the contribution from polarization modification (compared with unpolarized case) is proportional to polarization 4-vector, which is showed in Eq. \eqref{plane wave case}.}
Thus we can reconstruct integral in Eq. \eqref{vortex muon decay modification 2} as integral of linear terms of $\varphi_p$-dependent weight factors ($a_p$, $b_p$ and $c_p$), as showed in Eq. \eqref{cos and sin integral}.
\item
Spectral-azimuthal distribution of plane wave muon decay has azimuthal dependence described by trigonometric functions with frequencies $1$ and $2$ if 3-momentum of the muon is not parallel to $z$ axis for both unpolarized and polarized cases, which is discussed at last paragraph of Section \ref{section three} Part A.
If we rotate the 3-momentum of the plane wave with certain angle around $z$ axis, the azimuthal dependent distribtuion will rotate simultaneously with the same angle. But the shape of the distribution is not changed.
\end{itemize}
With the three prerequisites, we can understand contribution from unpolarized part and polarization modification part of vortex muon decay respectively.
Sum of them is showed in Eq. \eqref{vortex muon decay} and the modification part is showed in Eq. \eqref{vortex muon decay modification}.
For angular average of unpolarized part, we have no $\varphi_p$-dependent weight factors and simply get equal weight superposition of azimuthal distibutions that rotate along with rotating of plane wave 3-momentum.
The rotation scans a full circle of 3-momentum since momentum distribution of vortex state is just a circle.
Thus the result of the superposition is axisymetric, which means it exhibits any $Z_n$ symmetry.
For angular average of polarization modification part, we have additional $\varphi_p$-dependent weight factors for the superposition.
The distribution is modified by the weight factors which can be expanded as trigonometric series.
If $0<E_e<E_{e1}$, the trigonometric functions in the plane wave components distributions work as filters to trigonometric function components in the weight factors.
Thus, only frequencies $1$ (no $Z_n$ symmetry) and $2$ ($Z_2$) in trigonometric function components of the weight factors can survive in the result.
The distribution will be axisymetric if both of the frequencies do not exist.
If $E_{e1}<E_e<E_{e2}$, the trigonometric functions in the plane wave components distributions work as phase and amplitude modulators to trigonometric function components of the weight factors.
And the frequencies in trigonometric function components of the weight factors are not changed, which means the $Z_n$ symmetry of the weight factors is kept.
In summary, we get the result discussed in previous paragraphs and showed in Fig. \ref{electron azimuthal distribution z}.
The same logic applies to other interactions as long as the three prerequisties are satisfied.
Note that the frequencies in the third prerequisite can be different.

	\section{conclusions}\label{section five}
$Z_n$ symmetry is a discrete symmetry that is not taken into acount in current particle physics, since the plane wave approximation used to be postulated and can give nice result that agree well with experiment datas in most cases.
But the non-plane-wave nature of particles has some non-negligible effects  to collision of particles.
For example, the so called ``beam-size effect'' is verified in experiments at the MD-1 detector on the VEPP-4 collider, Novosibirsk 1981 \cite{Large impact parameter cut-off in the process e+e- to e+e-gamma}, which gives significant inconformity to standard QED calculation with plane wave approximation.
Calculation of the same process with Gaussian type states get quite reasonable agreement with the experiment data \cite{Deviation from standard QED at large distances: Influence of transverse dimensions of colliding
beams on bremsstrahlung}.
By considering the non-plane-wave particle states, we can imagine particles exhibiting $Z_n$ symmetry.
How does this kind of symmetry behave after particle interaction remains to be unknown.

In this paper, we select a series of peculiar non-plane-wave fermion states which we call them ``vortex fermion states with $Z_n$ symmetry'' to study behavior of $Z_n$ symmetry in a simple example of muon decay.
$Z_n$ symmetry of this kind of states are constructed in  momentum space and showed in coordinate space with some approximations.
They are just special cases of the so called ``spin-orbit states'' which are extensively researched for recent years.
We calculated the plane wave electron spectral-azimuthal distribution at certain emission angle after decay of muon in such state, which is only thing we can detect for muon decay.
The result shows that the $Z_n$ symmetry of initial state remains invariant after the decay process.
Concretely, it is verified by the fact that electron distribution of final state displays the same $Z_n$ symmetry as the initial vortex muon state do.
Other interactions that share some similar features with the process we  have considered in the paper may get the same result.
Three selection rules are listed to specify these cases, though they can only cover a small part of all the interactions.
So, the result may not represent general behavior of $Z_n$ symmetry in particle interactions.

Though there is no realization of such kind of muon states nowadays, their photon version have been produced. And generation methods of electron and neutron versions have been proposed by some researchers. The methods should be generalized to muon and other fermions.
The kind of vortex fermions provides a good stage for study $Z_n$ symmetry in  particle interactions.
We hope that the study here can stimulate the relavant research both  theoretically and experimentally.

	\section*{Acknowledgments}
	
	We thank the referee for careful reading and suggestions.
This work was supported by grants of the National Natural
Science Foundation of China (Grant No. 11975320) and the
Fundamental Research Funds for the Central Universities,
Sun Yat-sen University.
	
\section*{Appendix A: Relations between fermion polarization parameters}
We have two groups of parameters to describe polarization of plane wave fermions, i.e. $\alpha(\varphi_p),\,\delta(\varphi_p)$ for spinor in Eq.\eqref{spinor} and $a_p,\,b_p,\,c_p$ for polarization 4-vector in Eq.\eqref{jihuafenjie}.
They can be related to each other with one-to-one correspondence by three equations:
\begin{eqnarray}
&&\cos {\alpha(\varphi_p)}=b_p( \frac{1}{\gamma}\cos^2 \theta_0+\sin^2 \theta_0)+a_p\sin \theta_0 \cos \theta_0(\frac{1}{\gamma}-1),\label{eqn1}\\
&&\sin (\varphi_p-\delta(\varphi_p))= \frac{c_p}{\sin {\alpha (\varphi _p)}}, \label{eqn2}\\
&&\cos (\varphi_p-\delta(\varphi_p))= \frac{a_p\cos {\theta_0}-b_p\sin \theta_0 +\cos {\alpha (\varphi _p)} \sin \theta_0}{\sin {\alpha (\varphi _p)} \,\cos \theta_0}, \label{eqn3}
\end{eqnarray}
where $0<\alpha(\varphi_p)<\pi$ and $0<\delta(\varphi_p)<2\pi$.
It can be got by short calculation according to definition of the two different polarization descriptions.
For spinor case, combination of the two free angles $(\alpha,\,\delta)$ gives spin direction $\vec s_0$ of plane wave state at its rest frame.
For 4-vector case, $a_p,\,b_p$ and $c_p$ perform as weight factors of three orthogonal basis vectors at the moving frame, in which the 3-momentum of the state is $\bf p$.
Detail calculation is showed as follows.

\begin{figure}[!h]
		\centering
\includegraphics[width=0.9\textwidth]{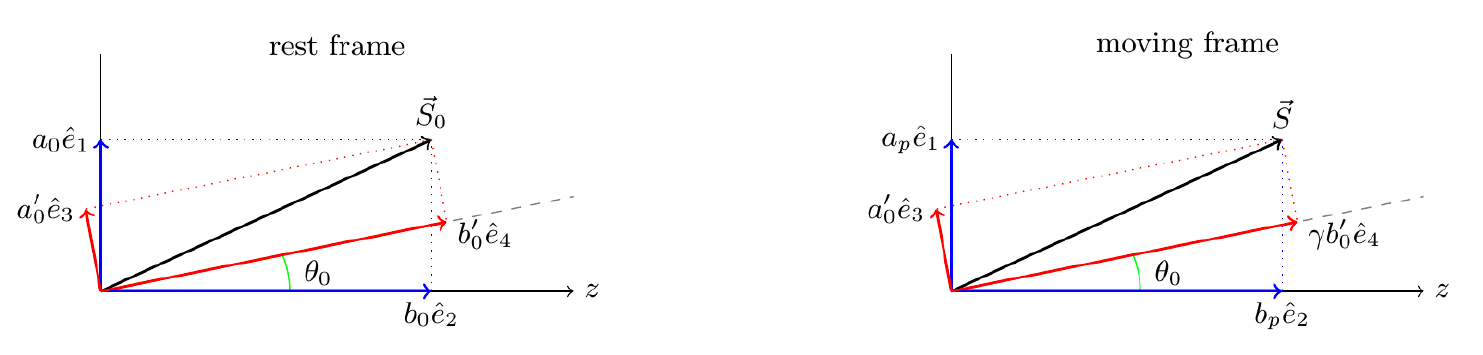}
		\caption{Polarization vector decomposition with different  unit vectors (labeled as $\hat e_1,\,\hat e_2,\,\hat e_3$ and $\hat e_4$) in different reference systems. The plane we choose includes both ${\bf p}$ and $z$ axis. $\vec S_0$ and $\vec S$ are projections of polarization vectors in the plane. Azimuthal component is vertical to the plane (not drawn) and do not change under boosting ({ {i.e.}} azimuthal polarization projection keeps unchanged: $c_0=c_p$) . $\theta_0$ corresponds to conical angle of vortex state.}\label{coordinates}
	\end{figure}

Under boosting from rest frame to moving frame with momentum $p^{\mu}$, polarization 4-vector is changed from $s_0^{\nu}=(0,\,{\bf S}_0)=(0,a_0,b_0,c_0)$ to $s^{\nu}=(S^0,\,{\bf S})=(S^0,a_p,b_p,c_p)$. For space part, only components that are parallel to the moving direction is changed (multiply by $\gamma$).
We use two figures in Fig. \ref{coordinates} to calculate relations in the two frames.
In the rest frame, we have 
\begin{eqnarray}
&&{\vec S}_0=a_0\hat e_1+b_0\hat e_2=a_0^{\prime}\hat e_3+b_0^{\prime}\hat e_4,\\
&&a_0=a_0^{\prime}\cos \theta_0+b_0^{\prime}\sin \theta_0,\\
&&b_0=b_0^{\prime}\cos \theta_0-a_0^{\prime}\sin \theta_0.
\end{eqnarray}
In the moving frame, we have
\begin{eqnarray}
&&{\vec S}=a_p\hat e_1+b_p\hat e_2=a_p^{\prime}\hat e_3+\gamma b_p^{\prime}\hat e_4,\\
&&a_p=a_0^{\prime}\cos \theta_0+\gamma b_0^{\prime}\sin \theta_0,\\
&&b_p=\gamma b_0^{\prime}\cos \theta_0-a_0^{\prime}\sin \theta_0.
\end{eqnarray}
We use angles to express unit vectors in different direction: ${\bf S}_0\rightarrow (\alpha (\varphi _p), \delta (\varphi _p)),\, \hat e_1\rightarrow (\pi/2, \varphi _p),\, \hat e_2\rightarrow (0, \varphi _p),\, \hat e_3\rightarrow (\theta _0+\pi/2, \varphi _p),\, \hat e_4\rightarrow (\theta_0, \varphi _p),\, \hat e_a\rightarrow (\pi/2, \varphi _p-\pi/2)$.
Here $\hat e_a=\hat e_1\times \hat e_2=\hat e_3\times \hat e_4$.
Since 
\begin{eqnarray}
&&{\bf S}_0\cdot \hat e_2=\cos \alpha (\varphi_p), \\
&&{\bf S}_0\cdot \hat e_1=a_0=a_p,\\
&&{\bf S}_0\cdot \hat e_2=b_0,\\
&&{\bf S}_0\cdot \hat e_a=c_p,
\end{eqnarray}
we can directly get Eq.\eqref{eqn1},
 and then Eq.\eqref{eqn2} and Eq.\eqref{eqn3} with all the equations given above.
{Polarization vector in the rest frame is normalized as $s_0^2=-a_0^2-b_0^2-c_0^2=-1$, which is Lorentz invariant.
Thus, in the moving frame, we can get the normalization condition, i.e. Eq.~\eqref{abceq}, by calculation with equations given above.}

\section*{Appendix B: Important integrals used in section \ref{section four}}
In section \ref{section four}, we discussed result of integral showed in Eq. \eqref{cos and sin integral}.
We do not give acurate result but just give general features of different terms in the equation.
All the discussions are based on some special integrals about trigonometric functions listed as follows.
For any integers $n$ and $n^{\prime}$, if they are not equal, we have
\begin{eqnarray}
\int ^{\phi+\tau}_{\phi-\tau}d\varphi_p\, \cos{{n^{\prime}}(\varphi_p-\phi)}\,\cos{n\varphi_p}=\frac{2({n^{\prime}}\cos{n\tau}\,\sin{{n^{\prime}}\tau}-n\sin{n\tau}\,\cos{{n^{\prime}}\tau})}{{n^{\prime}}^2-n^2}\cos{n\phi};\\
\int ^{\phi+\tau}_{\phi-\tau}d\varphi_p\, \cos{{n^{\prime}}(\varphi_p-\phi)}\,\sin{n\varphi_p}=\frac{2({n^{\prime}}\cos{n\tau}\,\sin{{n^{\prime}}\tau}-n\sin{n\tau}\,\cos{{n^{\prime}}\tau})}{{n^{\prime}}^2-n^2}\sin{n\phi};\\
\int ^{\phi+\tau}_{\phi-\tau}d\varphi_p\, \sin{{n^{\prime}}(\varphi_p-\phi)}\,\cos{n\varphi_p}=\frac{2({n^{\prime}}\sin{n\tau}\,\cos{{n^{\prime}}\tau}-n\cos{n\tau}\,\sin{{n^{\prime}}\tau})}{{n^{\prime}}^2-n^2}\sin{n\phi};\\
\int ^{\phi+\tau}_{\phi-\tau}d\varphi_p\, \sin{{n^{\prime}}(\varphi_p-\phi)}\,\sin{n\varphi_p}=\frac{2({n^{\prime}}\sin{n\tau}\,\cos{{n^{\prime}}\tau}-n\cos{n\tau}\,\sin{{n^{\prime}}\tau})}{n^2-{n^{\prime}}^2}\cos{n\phi}.
\end{eqnarray}
If they are equal, we have
\begin{eqnarray}
\int ^{\phi+\tau}_{\phi-\tau}d\varphi_p\, \cos{n(\varphi_p-\phi)}\,\cos{n\varphi_p}&=&\frac{2n\tau+\sin{2n\tau}}{2n}\cos{n\phi};\\
\int ^{\phi+\tau}_{\phi-\tau}d\varphi_p\, \cos{n(\varphi_p-\phi)}\,\sin{n\varphi_p}&=&\frac{2n\tau+\sin{2n\tau}}{2n}\sin{n\phi};\\
\int ^{\phi+\tau}_{\phi-\tau}d\varphi_p\, \sin{n(\varphi_p-\phi)}\,\cos{n\varphi_p}&=&\frac{-2n\tau+\sin{2n\tau}}{2n}\sin{n\phi};\\
\int ^{\phi+\tau}_{\phi-\tau}d\varphi_p\, \sin{n(\varphi_p-\phi)}\,\sin{n\varphi_p}&=&\frac{2n\tau-\sin{2n\tau}}{2n}\cos{n\phi}.
\end{eqnarray}
When $\tau=\pi$, we just get the orthogonal theorem for integral of trigonometric functions.

\end{document}